\documentclass[12pt]{article}
\renewcommand{\baselinestretch}{1.2}
\usepackage{indentfirst}
\usepackage{amsmath}
\usepackage{amssymb}
\usepackage{amsfonts}
\usepackage{amscd}
\usepackage{amsbsy}
\usepackage{amsthm}
\usepackage{latexsym}
\usepackage{graphicx,color} 
\usepackage[dvipsnames]{xcolor}
\usepackage[colorlinks]{hyperref}
\hypersetup{linkcolor=blue,citecolor=blue,urlcolor=blue}
\def\nn{\nonumber}       
\def\beq{\begin{eqnarray}}
\def\eeq{\end{eqnarray}}
\def\ln{\,\mbox{ln}\,}

\DeclareMathOperator{\cx}{\square}

\def\al{\alpha}
\def\be{\beta}
\def\ch{\chi}

\def\de{\delta}

\def\ze{\zeta}

\def\pa{\partial}

\def\rh{\rho}
\def\si{\sigma}

\def\ph{\varphi}
\def\ta{\tau}
\def\th{\theta}

\def\Ga{\Gamma}
\def\De{\Delta}
\def\La{\Lambda}

\usepackage[symbol]{footmisc} 
\usepackage{float} 
\usepackage{cite}  
\usepackage{titlesec}

\usepackage{ulem} 

\titleformat*{\section}{\large\bfseries}
\titleformat*{\subsection}{\normalsize\bfseries}
\begin{document}

\begin{center}
\renewcommand*{\thefootnote}{\fnsymbol{footnote}}
{\Large 
Low-energy limit in the anomaly-induced action
\\
and the semiclassical cosmological bounce}
\vskip 4mm

{\bf Wagno Cesar e Silva}\,
\,$^{a}$
\hspace{-1mm}\footnote{E-mail address: \ wagnorion@gmail.com},
\quad
{\bf Nicolas R. Bertini}\,
\,$^{b}$
\hspace{-1mm}\footnote{E-mail address: \ nicolas.bertini@hotmail.com}
\quad
and
\quad
{\bf Ilya L. Shapiro}\,
\,$^{b}$
\hspace{-1mm}\footnote{E-mail address: \ ilyashapiro2003@ufjf.br}
\vskip 4mm

{
${a)}$ Centro Brasileiro de Pesquisas F\'{\i}sicas,
\\
Rua Dr. Xavier Sigaud 150, Urca, 22290-180, Rio de Janeiro, RJ, Brazil
\vskip 2mm

${b)}$ Departamento de F\'{\i}sica, ICE, Universidade Federal de Juiz de Fora,
\\
Campus Universitário, 36036-900, Juiz de Fora, MG, Brazil
}

\end{center}
\vskip 2mm


\begin{quotation}

\centerline{\large\bf Abstract}
\vskip 1mm

\noindent
In the recently proposed scenario, the cosmological bounce occurs
because the initially contracting Universe is not empty. In the
region close to singularity, matter contents of the Universe heat
up and effectively become radiation. Then, the trace anomaly
automatically provides bounce if the overall beta function in the
matter sector is positive. Independent of the remaining open
questions on the quantum field theory side, it is interesting to
consider this model from the cosmological perspective. In the
present work, we develop the general formalism which is a necessary
step for exploring the primordial cosmological perturbations. The
main technical development is the formulation of the low-energy
version for the nonlocal part of the effective action. The complete
form of this action can be done local using two auxiliary scalars.
In our new version, there are more scalars, but this enables one to
avoid higher derivatives.
\vskip 3mm



\end{quotation}
	
\setcounter{footnote}{0} 
\renewcommand*{\thefootnote}{\arabic{footnote}}

\section{Introduction}
\label{sec1}

The trace anomaly and the anomaly-induced effective action
form important parts of the semiclassical gravity, i.e., the theory
of quantum matter fields on a classical gravitational background.
One of the typical features of  the anomaly-induced action is the
presence of higher derivatives. The corresponding terms do not
produce ghosts in the spectrum on a flat spacetime, but the
situation may change if the  background metric is nontrivial. In any
case, the ghosts become visible only at the energies comparable
to the Planck scale. Thus, it is tempting to formulate an effective
approach and separate that part of the anomaly-induced action
which remains relevant in the low-energy domain (IR), that is
well below the Planck scale. In this work, we elaborate such
a formulation and apply it to the bounce solution in cosmology.
One of the purposes is to check the consistency of our approach.

It is a well-known  that the initial singularity represents one
of the most challenging theoretical problems of the standard
cosmological paradigm. To address this issue, several main approaches
have emerged,  including the bouncing models (see, e.g., the reviews
\cite{Novello2008,Peter2015}), and emergent universe scenarios
\cite{Ellis2002,Ellis2004}. Non-singular bouncing models provide
a consistent framework where the initial singularity is resolved by
a smooth transition connecting a contracting pre-bounce phase to
the current expanding phase. At the bounce point, the scale factor
of the Universe is reaching a finite, strictly positive minimum.

In the context of trace anomaly \cite{CapDuf-74,duff77}, the
possibility to achieve bouncing solutions was considered in
Refs.~\cite{star,Anderson83-84}, and discussed  using anomaly-induced
effective action in \cite{AnJu}. In both cases, the bounce occurs
because of the purely gravitational terms in anomaly, and the typical
energy density in the bounce point has the Planck order of
magnitude. More recently, another possibility was explored by two
of the present authors in \cite{AnoBo21,BoRa24}. Assuming that the
initially contracting Universe has some matter contents, there is an
explicit analytical solution describing
the cosmological bounce without singularity.
The bounce occurs owing to the equilibrium between the classical
radiation term and the quantum correction in the radiation-gravitational
sector. The form of these loop contributions is well-known and does
not require anything besides the well-established results of quantum
field theory. If comparing with the previously known bouncing models
based on exotic matter fields, the anomaly-induced correction to
radiation plays the role of the phantom scalar \cite{BoNePa02}.

The bounce solution  \cite{BoRa24} does not require any sort
of \textit{ad hoc} assumptions, i.e., modifying the action of gravity,
introducing a scalar field, or accounting for the vacuum quantum
effects. However, this model has the following main problems:

\textit{i)} assuming the beta function of QED, the Hubble
parameter in the point of the bounce is in the deep transplanckian
area. This is not a forbidding point because the bounce is expected
at the energy scale between inflation (i.e., above $10^{12}\,GeV$)
and the Planck scale, which is approximately $10^{19}\,GeV$. At
these energies, not only QED, but even the Minimal Standard
Model (MSM) is not assumed to describe particle Physics, hence
the described situation indicates to some special version of Physics
beyond the MSM, with stronger interactions and the corresponding
scheme of decoupling of extra degrees of freedom.

\textit{ii)} The consistent model of bounce imposes certain
requirements to the cosmological perturbations and stability
\cite{Novello2008,Peter2015}. On the other hand, the study of
stability in the model based on the anomaly-induced action may
be complicated because of the presence of higher derivative terms
in the gravitational sector. However, since we intend to use this
model at the energies below the Planck scale, it can be expected
that the higher derivative ghosts will not be generated
\cite{HD-Stab}, which is consistent with the traditional
effective approach \cite{Simon-90}. The best way to put this
approach into practise is rewriting the low-energy limit of the
anomaly-induced action in a second-order form using new
auxiliary scalars. The correctness test of this reformulation should
be the existence of the bounce solution found in \cite{BoRa24}.

The described procedure is the subject of the present work.
We obtain a new covariant local formulation of the
anomaly-induced action, which can be regarded as a low-energy
(with respect to the Planck scale) approximation to the well-known
general expression \cite{rie,frts84,a} (see also \cite{OUP} for the
modern simplified derivation, introduction, and further references).
The new form of induced action should be a suitable framework for
analyzing cosmological perturbations, but it can also be used for
other purposes.

It is important to note that, in the bounce model under discussion,
the perturbations should be analyzed on the basis of a covariant
local representation of induced action. The local version of this
action includes two auxiliary scalar fields \cite{a,MaMo}.
According to the criterion of \cite{Peter2015}, there are chances
for a consistent bounce model, but the absence or presence of
possible pathologies depends on the dynamics of the perturbations.
As in other bounce models, the study of stability at the point of
transition between the expansion and contraction phases of the
universe is a typically complicated issue that requires an analysis
at the level of cosmological perturbations. It is known that, in
the vicinity of the bounce, the time derivative $\dot{H}$ is
necessarily positive and this implies the violation of the null energy
condition (NEC), $\rh+p \geq 0$. This feature may lead to instabilities
in cosmological perturbations \cite{Visser1998,Novello2008} (see
also \cite{PatrickNelson01} for an alternative discussion).
Indeed, the healthy violation of the NEC, i.e., without generating
catastrophic instabilities, is a major challenge in the construction of
many bounce models.  However, there are strong indications in the
literature that the NEC violation by quantum corrections may not
lead to inconsistencies \cite{Ford2003} and that the same is true in
the theories with scalar fields \cite{Rubakov14,IjjStein16}. Both
arguments can be applied in the case of the model presented in
\cite{BoRa24}.

The paper is structured as follows. Section \ref{sec2} presents a
brief review of the anomaly-induced action, including the
contribution from the radiation sector, and of the bounce
solution of Ref.~\cite{BoRa24}. In Sec~\ref{sec2.1} we introduce
a new formulation of the low-energy version of the anomaly-induced
action \cite{GianMott09} in terms of auxiliary fields.
In Sect.~\ref{sec3}, we derive the background equations of motion,
establish the analytical conditions for the existence of nonsingular
bouncing solutions, and investigate the cosmological dynamics
numerically for two different classes of initial conditions.
Finally, in Sect.~\ref{sec6}, we draw our conclusions.

We adopt the natural units with $c=\hbar=1$,
the metric signature $(+,-,-,-)$, and the definition of the
curvature tensor
$R^\al_{\,\,\,\be\mu\nu} = \pa_\mu \Ga^\al_{\be\nu}+ ...$.

\section{Anomaly-induced action with radiation}
\label{sec2}

As we discussed in the Introduction, in the contracting Universe,
at some point all matter contents becomes radiation, i.e., can be
regarded a set of massless fields. At classical level, this means
that all fields are conformal, with the vanishing trace of the
energy-momentum tensor. In this situation, the leading one-loop
quantum contributions correspond to the conformal anomaly
\cite{CapDuf-74,duff77}. For the sake of simplicity, one can trade
all matter fields to electromagnetic or Yang-Mills radiation
described by the potential  $A_\mu$, and then we get
\cite{RadiAna,BoRa24}
\beq
\langle \mathcal{T} \rangle
& = & -\,\frac{2}{\sqrt{-g}}\,g_{\alpha\beta}\,
\frac{\de \Ga}{\de g_{\alpha\beta}}
\,\,=\,\,-\,Y(g_{\mu\nu},A_{\mu}) - bE_4 - c\,\square R\,,
\label{anomaly}
\eeq
where $Y(g_{\mu\nu},A_{\mu})=w C^2-\frac14\be g^2F^2$
represents the conformal invariant terms. In this expression,
$\,F^2=F_{\mu\nu}F^{\mu\nu}\,$ is the square of the gauge field
strength and $C^2$ is the square of the Weyl tensor. Furthermore,
$E_4= R_{\mu\nu\al\be}^2 - 4 R_{\al\be}^2 + R^2$ is the
Gauss-Bonnet invariant and, finally, $\Ga$ is the one-loop
renormalized effective action.

Using the conformal parametrization of the metric,
\beq
g_{\al\be}\,=\,{\bar g}_{\al\be}\,e^{2\si}\,,
\label{confmet}
\eeq
one can write the anomalous trace as
\beq
\langle \mathcal{T} \rangle
& = &
- \,\frac{1}{\sqrt{-{\bar g}}}\,
\frac{\de\,\Ga[{\bar g}_{\al\be}\,e^{2\si}]}{\de \si}
\,\bigg|_{{\bar g_{\al\be}}\rightarrow g_{\alpha\beta}, \,\,
\si\rightarrow 0}\,\,\,.
\label{anomaly-sig}
\eeq

The coefficient of the $\,F^2$-term depends on the beta function,
which is conveniently presented in the form $\be g^4$, where
\beq
\be \,=\,-\,
\frac{2}{(4\pi)^2}\,\Big(
\frac{11}{3}\,C_1 - \frac{1}{6}\,N_{cs}
- \frac{4}{3}\,N_f\Big).
\label{ym19}
\eeq
Here $N_{cs}$ and $N_f$ are the numbers of complex scalars and
fermions coupled to the given vector field. $C_1$ is the Casimir
operator of the gauge group, which is zero in the Abelian theory.
In the non-Abelian case, $C_1$ is positive, providing the asymptotic
freedom \cite{GrossWilczek,Politzer}.

Using Eq.~(\ref{anomaly}), one can find a solution to the effective
action. In the non-covariant local form \cite{rie,frts84} and the
parametrization (\ref{confmet}), this solution has the form
\beq
&&
\Ga_{\textrm{ind}}\,=\,
S_c[{\bar g}_{\mu\nu}]
\,+\,\int d^4 x \sqrt{-{\bar g}}
\,\bigg\{\si  \bar{Y}
+\, b\si \Big({\bar E}
-\frac23 {\bar \cx}
{\bar R}\Big)
+ 2b\si{\bar \De}_4\si\bigg\}
\nn
\\
&&
\qquad
\quad
\,-\,\frac{2 b+3c}{36}\int d^4 x\sqrt{-g}\,R^2,
\label{EA_induced}
\eeq
where $S_c$ is an arbitrary conformal functional and
all quantities with bars are constructed with the fiducial
metric $\bar g_{\al\be}$ and
$ \Delta_4=e^{-4\si}\bar{\Delta}_4$ is the Paneitz operator
\cite{Paneitz},
\beq
\Delta_4\,=\,
\cx^2
+ 2R^{\mu\nu}\nabla_{\mu}\nabla_{\nu}
- \dfrac{2}{3}R\square
+\dfrac{1}{3}(\nabla^{\mu}R)\nabla_{\mu}.
\label{Paneitz_op}
\eeq
We note that the integration constant
$S_c$ in Eq.\,\eqref{EA_induced} can be safely neglected in
zeroth-order cosmology, since this term does not influence the
dynamics of the conformal factor.

The non-local covariant solution of \eqref{anomaly} is derived using
the conformally covariant Green function of the Paneitz operator
$\De_4$, resulting in (see, e.g., \cite{OUP} for the details)
\beq
&&
\Ga_{\textrm{ind}}\,=
S_c
+\frac{b}{8}\int_{x}\int_{y}
\Big(E_4 - \frac23\cx R\Big)_x
\,G(x,y)\,\Big(E_4 - \frac23 \cx R\Big)_y
\nn
\\
&&
\qquad
\quad
+\frac{1}{4}\int_{x}\int_{y}
Y(x)
\,G(x,y)\,\Big(E_4 - \frac23\cx R\Big)_y
-\frac{2 b+3c}{36}\int_x R^2,
\label{Gar}
\eeq
with the compact notations for the integrals
$\int_x \equiv \int d^4x\,\sqrt{-g(x)}$. $\,G(x,y)\,$ are the
two Green functions of the same operator (\ref{Paneitz_op}).
These Green functions may differ by the
choice of boundary conditions, regardless responding
the same differential operator.

Besides the representations \eqref{Gar} and \eqref{EA_induced},
one can formulate the covariant local form for this action,
by introducing two auxiliary scalar fields \cite{a}. This is the
most useful formulation for applications, such as classification
of vacuum states in the vicinity of a black hole \cite{balsan}, or
the reaction
of gravitational waves to the presence of higher derivatives
\cite{AnJu-wave}. Introducing the auxiliary scalars enables one to
trade imposing the boundary conditions for the Green functions
to the same conditions for the scalar fields, that proved more
useful for the mentioned applications.
Another possibility related to the nonlocal expression  (\ref{Gar})
is to consider its low-energy limit (IR). 
This procedure provides certain
simplifications, as originally found in \cite{MotVaul,GianMott09}
for the metric-electromagnetic background and then extended to
the scalar field in \cite{AnoInt23}. This approach can be also
adapted by introducing auxiliary scalar fields, as we discuss in
the next section.

The bounce model of Ref.~\cite{BoRa24} is based on 
the non-covariant local representation \eqref{EA_induced}, in which 
the anomaly-induced contribution is dominated by the radiation 
sector. In this case, the relevant part of effective action is
\beq
\Ga \,=\, -\,\frac{1}{16\pi G}\int d^4x\sqrt{-g}\,\big(R+2\La\big)
\,-\, \frac{\,\be g^2}{4\,}  \int d^4x\,\sqrt{-\bar{g}}\,\bar{F}^2\si,
\label{Garel}
\eeq
where 
$\bar{F}^2 = \bar{g}^{\mu\al} \bar{g}^{\nu\be} F_{\mu\nu}F_{\al\be}$.
In the present work, we are interested in a covariant formulation of 
\eqref{Garel}. For this purpose, it is more convenient to start our 
construction from the representation \eqref{Gar}.



\section{Low energy regime and covariant local representation}
\label{sec2.1}

The procedure we adopt here consists of the two steps: $i)$ taking
the low-energy limit of the covariant and non-local representation
\eqref{Gar}, and $ii)$ rewriting the non-localities through the
introduction of auxiliary fields.
In the first part, let us use the scheme analogous to
Refs.~\cite{MotVaul,GianMott09,AnoInt23}.
We assume the weak-curvature limit, i.e., that the radiation sector
dominates over the curvature terms. This means,
\beq
|\square R| \gg |R^2_{....}|
\quad
\textrm{and}
\quad
|F^2| \gg |R^2_{....}|.
\label{Approximations}
\eeq
Then, the non-local structures in the induced action
\eqref{Gar} simplify owing to
\beq
G=\Delta_4^{-1} \approx \square^{-2}\,.
\label{Green_function}
\eeq
Thus, the leading terms are those that contain $F^2$ and are linear
in curvature, i.e.,
\beq
&&
\Ga_{\textrm{ind}}^{\textrm{IR}}
\,\,\approx\,\,
\frac{\be g^2}{24}\int_{x}\int_{y}
F^2(x)
\,\Big(\frac{1}{\square^2}\Big)_{x,y}\,(\square R)_y\,\,.
\label{EA_red}
\eeq

Next, following the approach of \cite{a}, we make use of the
auxiliary field description to parameterize the non-localities
present in \eqref{EA_red}. An intermediate step is rewriting the
integrand in \eqref{EA_red} as a sum of two Gaussian terms,
\beq
F^2\,\frac{1}{\square^2} \,\square R
&=&
\frac14\bigg[\big(F^2+\square R\big)
\frac{1}{\square^2}
\big(F^2+\square R\big)
-\big(F^2-\square R\big)
\frac{1}{\square^2}
\big(F^2-\square R\big)\bigg].
\label{manipulation}
\eeq
The local representation is achieved by introducing
two auxiliary scalar fields
\beq
&&
\phi(x)
\,=\,
\int_y \Big(\frac{1}{\square^2}\Big)_{x,y}\,(F^2+\square R)_y\,,
\nn
\\
&&
\psi(x)
\,=\,
\int_y \Big(\frac{1}{\square^2}\Big)_{x,y}\,(F^2-\square R)_y\,.
\label{aux_fields}
\eeq
Using these fields,
the nonlocal covariant part of the anomaly-induced action, together
with the Einstein-Hilbert term and the cosmological constant, can be
cast in the classically equivalent form
\beq
&&
\Ga_{\textrm{IR}}
\,=\,
-\,\frac{1}{16\pi G}\int_x\,\big(R+2\La\big)
+\frac{\be g^2}{48}\, \int_x\,\bigg\{
-\frac12(\square\ch)\,\square\ph
+R\,\square\ch
+F^2 \ph
\bigg\},
\label{Action}
\eeq
where we employed the field redefinitions $\ph=\phi-\psi$ and
$\ch=\phi+\psi$ to express the anomalous part in a more compact
form:
\beq
&&
\ph(x) \,=\, \phi(x) - \psi(x)
\,=\,
\int_y \Big(\frac{2}{\square^2}\Big)_{x,y}\,(\square R)_y
\,=\,
\int_y \Big(\frac{2}{\square}\Big)_{x,y}\,R(y)\,,
\nn
\\
&&
\chi(x) \,=\, \phi(x) + \psi(x)
\,=\,
\int_y \Big(\frac{2}{\square^2}\Big)_{x,y}\,(F^2)_y\,.
\label{auxphchi}
\eeq

The action \eqref{Action} still contains higher-derivative terms,
including fourth-order contributions in the kinetic sector of the
scalar fields, such as $(\square\ch)\,\square\ph$.
For practical purposes, including the analysis of cosmic
perturbations, it is appropriate to recast it in a form that involves
only second-order derivatives. For this, we apply more auxiliary
fields. Introducing two Lagrange multipliers, $ \ze $
and $\xi$, and defining the auxiliary fields $ \th=\square\ph $ and
$ \ta=\square\ch $, the anomalous part of \eqref{Action} becomes
\beq
\Ga_{\textrm{ind}}^{\textrm{IR}}[\th,\ta,\ph,\ch]
\,\,=\,\,
\frac{\be g^2}{48}\int_{x}
\bigg\{
-\frac{1}{2}\ta\th
+R\ta
+F^2\ph
+\ze\big(\th-\square\ph\big)
+\xi\big(\ta-\square\ch\big)
\bigg\}.
\label{scalar_action2}
\eeq

The variation with respect to $\th$ and $\ta$ yields two algebraic
constraint equations
\beq
&&
-\,\frac12\ta+\ze \,=\,0,
\quad \mbox{and} \quad
-\,\frac12\th+R+\xi\,=\,0.
\label{constr_eq}
\eeq
Substituting \eqref{constr_eq} into \eqref{scalar_action2} and
including the EH sector, the action \eqref{Action} takes the form
\beq
\label{Action_local}
&&
\Ga_{\textrm{IR}}
\,\,=\,\,
-\,\frac{1}{16\pi G}\int_x\,\big(R+2\La\big)
-\frac{\be g^2}{48}\, \int_x\,\Big\{
\ph\square\ze
+\ch \square \xi
-2\ze(R+\xi)
-F^2\ph
\Big\}.
\qquad
\eeq
One can note that scalar fields $ \ph $ and $\ch$ are dimensionless,
i.e., $ [\ph]=[\ch]=0 $, and should be assumed conformally invariant.
In contrast, the fields $\ze$ and $\xi$ have canonical mass
dimensions \ $[\ze] = [\xi] = 2$, and transform under a conformal
rescaling (\ref{confmet}) according to $\ze = e^{-2\si} \bar{\ze}$
and $\xi = e^{-2\si} \bar{\xi}$, respectively.

The effective action \eqref{Action_local} represents a covariant
formulation of the model with anomaly-induced corrections in
the IR, that is, in the region below the Planck scale. The next step
is to prepare the background for recovering the bounce solution
of \cite{BoRa24} using the new action. However, more
transformations are in order before we start this part.

To recover the action \eqref{EA_red} from \eqref{Action_local}, one
has to express the new auxiliary fields in terms of the corresponding
non-local structures involving $R$ and $F^2$. Using the definitions
(\ref{auxphchi}), the new auxiliary fields $ \th=\square\ph $,
$\ta=\square\ch$, and the constraint equations \eqref{constr_eq},
we arrive at the relatively simple identifications
\beq
&&
\th(x) \,=\, \cx_x \ph(x)
\,=\,
2 \int_y \cx_x \Big(\frac{1}{\cx}\Big)_{x,y}\,R_y
\,=\, 2 R_x\,,
\nn
\\
&&
\xi(x) \,=\, \frac12 \th(x) - R_x \,=\, 0\,,
\nn
\\
&&
\ze(x) \,=\, \frac12 \tau(x)
\,=\, \frac12 \cx_x \chi(x)
\,=\, \int_y \Big(\frac{1}{\cx}\Big)_{x,y}\,F^2_y\,.
\label{aux_xize}
\eeq
According to \eqref{aux_xize},
the field $\xi$ does not carry physical degrees of freedom.
Consequently, all non-local information is encoded in the auxiliary
fields $\ph$ and $\ze$.\footnote{
This property allows for further simplifications. In particular,
the constraint condition $\xi=0$ indicates that the field $\xi$ can
be eliminated from the action \eqref{Action_local}, leading to
an ``on-shell reduced'' formulation with only two auxiliary fields
$\ph$ and $\ze$. Nevertheless, it proves convenient to retain the
form \eqref{Action_local} in the numerical analysis of the
background cosmological solutions. 
}

The metric field equation derived from the action
\eqref{Action_local} is given by
\begin{align}
\big(1-2\kappa\zeta\big)G_{\mu\nu}
- \Lambda g_{\mu\nu}
\,=\,
\kappa T_{\mu\nu}^{(S)} + \kappa\varphi T_{\mu\nu}^{(F)},
\label{EH_eq}
\end{align}
where $G_{\mu\nu}$ is the Einstein tensor,
\ $\kappa = \beta g^2\pi/3M_P^2$, \ and
\beq
T_{\mu\nu}^{(S)}
&=& \frac{1}{2}\Big[ (\nabla_{\mu}\varphi) \nabla_{\nu}\zeta
+ (\nabla_{\nu}\varphi) \nabla_{\mu}\zeta
+ (\nabla_{\mu}\chi) \nabla_{\nu}\xi
+ (\nabla_{\nu}\chi) \nabla_{\mu}\xi
- 4\nabla_{\mu}\nabla_{\nu}\zeta \Big]
\nonumber
\\
&&
-\,\,
\frac12\,g_{\mu\nu}
\Big[  (\nabla_{\lambda}\xi) \nabla^{\lambda}\chi
+  (\nabla_{\lambda}\zeta)\nabla^{\lambda}\varphi
+ 2\zeta\xi -4\square \zeta \Big]\,,
\nonumber
\\
T_{\mu\nu}^{(F)}
&=&
2F_{\mu\alpha}F^{\alpha}_{~\;\nu}
- \frac{1}{2}\,g_{\mu\nu}F^2.
\eeq
On top of this, applying the variational principle with respect to
other fields, we arrive at the following covariant equations:
\beq
&&
\square\ze
-F^2
\,=\, 0,
\nn
\\
&&
\square\ph
-2\big(R+\xi\big)
\,=\, 0,
\nn
\\
&&
\square\xi
\,=\, 0,
\nn
\\
&&
\square\ch
-2\ze
\,=\, 0,\label{cov_eq}
\nn
\\
&&
\nabla_{\nu}(\varphi F_{\mu}^{\;~\nu})=0.
\label{eq:inc-rad}
\eeq

Now we have all the necessary ingredients to verify whether the
representation \eqref{Action_local} passes the main test, that means
it admits a nonsingular bouncing solution for the conformal factor.
This question will be addressed in the next section.

\section{Background cosmology}
\label{sec3}

We consider an isotropic and homogeneous background described
by the flat Friedmann-Lemaître-Robertson-Walker (FLRW) metric
\beq
ds^2
\,=\,
dt^2
- a^2(t) \big(dx^2+dy^2+dz^2\big),
\label{FLRW_metric}
\eeq
where $a(t)$ is the scale factor. The modified Friedmann equations
and the equations for the auxiliary fields obtained from
Eqs.~\eqref{EH_eq} -- \eqref{cov_eq} are
\beq
&&
(1-2\kappa\zeta)\big(3H^2+2\dot H\big)
+\frac{\kappa}{2}
\bigg[\frac{1}{3}
F^2\ph
+2 \xi\ze
+\dot \ch \dot \xi
-8 H \dot \ze
+\dot \ph \dot \ze
-4 \ddot \ze
\bigg]
-\La
\,=\,
0,
\qquad
\label{Friedmann_eq}
\\
&&
(1-2\kappa\zeta)3H^2
+\frac{\kappa}{2}
\Big[
F^2\ph
+2 \xi\ze
-\dot \ch \dot \xi
-12 H \dot \ze
-\dot \ph \dot \ze
\Big]
-\La
\,=\,0,
\qquad
\label{00_eq}
\\
&&
\ddot \ze
+3H \dot \ze
-F^2
\,=\, 0,\label{eq:zeta}
\\
&&
\ddot \ph
+3H \dot \ph
-2\big[\xi - 6\dot H -12H^2\big]
\,=\, 0,\label{eq:varphi}
\\
&&
\ddot \xi
+3 H \dot \xi
\,=\, 0,\label{eq:xi}
\\
&&
\ddot \ch
+3 H \dot \ch
-2\ze
\,=\, 0.
\label{eq:chi}
\eeq
The dot denotes derivatives with respect to cosmic time $t$,
and $H\equiv\dot{a}/a$ is the Hubble parameter. Eqs.~\eqref{00_eq}
and \eqref{Friedmann_eq} correspond to the $00$ and $ij$
components of \eqref{EH_eq}, respectively.

As can be seen in Appendix \ref{apdx}, Eq.~\eqref{eq:inc-rad}
can be solved analytically, yielding
\beq
F^2 = \frac{M^2}{a^{4}}-\frac{N^2}{\varphi^{2}a^{4}},
\label{eq:F2}
\eeq
where $M^{2}$ and $N^{2}$ are integration constants fixed by
the initial conditions.
Compared with the non-covariant model, Eq.~\eqref{Garel}, in
which the gauge-field invariant $\bar{F}^{2}$ is defined with the
fiducial metric, the present formulation admits a more general
expression for $F^{2}$, depending explicitly on the scale factor
$a(t)$ and the auxiliary field $\varphi(t)$.

To complement the verification of the equivalence between the action
\eqref{Action_local} and the non-covariant formulation employed in
Ref.~\cite{BoRa24}, the next step is to solve the system of equations
above. However, owing to its nonlinear and strongly coupled structure,
obtaining a closed-form analytical solution is highly nontrivial, if
possible at all. For the purposes of the present work, it suffices to
derive the conditions for bounce solutions and to solve the system
numerically under physically consistent initial conditions, which
fully determines the background cosmological dynamics.


\subsection{Bounce solution requirements}
\label{sec:bouce-conditions}

Let us consider the cosmological solution with the bounce at
time instant $t=t_b$. The scale factor and Hubble parameter have
to satisfy the restrictions
\begin{equation}
a(t_b) = a_b > 0,
\qquad
H(t_b)=0,
\qquad
\dot{H}(t_b) > 0\,.
\end{equation}
From here on, for any function $f(t)$, we use the notation
$f_b = f(t_b)$. For instance, $H_b = 0$. Other notations will be
introduced in what follows.

Assuming that the background solution is smooth in a
neighborhood of $t_b$, all dynamical variables admit a local
Taylor expansion in $\tau=t-t_b$. In particular, the scale factor
can be written as
\begin{align}
a(\tau) \,=\, a_b + \sum_{n= 2}^{\infty} a_n \tau^n,
\qquad
a_n \,\equiv\,
\frac{1}{n!}\frac{d^{n}a}{d\tau^{n}}\bigg|_{\tau = 0}\,.
\label{eq:a-expand}
\end{align}
A necessary local condition for the existence of a regular bounce
is $a_2 = \ddot{a}_b/2>0$.
Using the relation $\ddot{a}/a = H^{2} + \dot{H}$, we get
$\ddot{a}_b = a_b \dot{H}_b$. Using Eq.~\eqref{00_eq}
at the point $t_b$, we find
\begin{align}
\frac{\kappa}{2}\big( F^{2}_{b}\varphi_b
+ 2\xi_b\zeta_b
- \dot{\chi}_b \dot{\xi}_b
- \dot{\varphi}_b \dot{\zeta}_b \big)
\,=\,\Lambda .
\end{align}
Substituting this into \eqref{Friedmann_eq}, along with the
relation $\ddot{\zeta}_b = F^{2}_b$ obtained from \eqref{eq:zeta},
we get
\begin{align}
\left( 1- 2\kappa\zeta_b \right) \dot{H}_b
\,=\,
\frac{\kappa}{2}\Big[2F_{b}^{2} + \frac{1}{3}F^2_b\varphi_{b}
- \dot\chi_b \dot\xi_b - \dot\varphi_b \dot\zeta_b \Big].
\end{align}
This relation provides a local criterion for the sign of $\dot H_b$,
and consequently for the local fulfillment of the bounce condition.
For $1-2\kappa\zeta_b\neq0$, one obtains
\begin{align}
a_2
\,=\,
\frac{\kappa a_b}{4(1-2\kappa\zeta_b)}
\Big( 2F_{b}^{2} + \frac{1}{3}F^2_b\varphi_{b}
- \dot\chi_b \dot\xi_b - \dot\varphi_b \dot\zeta_b
\Big).
\label{eq:a_2general}
\end{align}

Motivated by the time-symmetric nonsingular solutions
\cite{BoRa24}, it is instructive to consider the time-symmetric
branch of the present localized covariant model. At the background
level, this branch is characterized by invariance under time
reflection around the bounce point, i.e., $a(t_b+\tau)=a(t_b-\tau)$.
Assuming that the anomalous sector shares the same symmetry, we
impose even parity for the auxiliary fields, which implies
\begin{align}
	\dot\zeta_b = \dot\varphi_b = \dot\xi_b = \dot\chi_b = 0 .
	\label{eq:dfields0}
\end{align}
Under these assumptions, the coefficient of the quadratic term
in Eq.~\eqref{eq:a-expand} reduces to
\begin{align}
a_2 =  \frac{\kappa a_b F^2_b}{	4\left( 1- 2\kappa \zeta_b \right)}
\left( 2+ \frac{\varphi_b}{3} \right).
\end{align}

The obtained relations play two roles in the subsequent analysis.
First, they provide analytical criteria that identify the
quantities governing the local realization of the bounce through
the coefficient $a_2$. On the other hand, they provide a consistency
check for the numerical solutions obtained below. The details of
the bounce are related to the values of auxiliary fields and of the
effective radiation term evaluated at the transition point.

\subsection{Initial conditions and numerical analysis}
\label{sec3.2}

In what follows, we explore two classes of initial conditions. In the
first case, the data are specified during the asymptotic contracting
phase, providing the most direct continuation of the non-local
formulation studied in Ref.~\cite{BoRa24}. In the second version,
the initial conditions are imposed directly at the bounce, where all
auxiliary fields are set to zero. This latter choice is not intended
to reproduce the non-local completion, but rather to test the
robustness of the bounce mechanism. If a nonsingular transition
still occurs in the absence of inherited auxiliary-field amplitudes,
it indicates that the bounce is a genuine dynamical feature of the
localized theory rather than an artifact of the asymptotic
construction.

\subsubsection{Initial conditions during the contraction phase}
\label{sec3.2.1}

Our first purpose is to construct initial conditions compatible with
the system of equations of motion. To this end, we assume that the
Universe initially evolves through a quasi-de Sitter contracting
phase, well before the nonsingular transition. During this stage,
the Hubble parameter is approximately constant and negative,
$H(t) \simeq H_c <0 $ and $|\dot{H}_c|\ll H_c^2$. Moreover,
since the initial hypersurface is chosen sufficiently far from the
bounce, the anomalous contribution proportional to
$N^{2}/(\varphi^{2}a^{4})$ is treated as a subleading correction
to the effective radiation sector. The full dynamics, including this
anomalous contribution, is subsequently recovered through the
numerical integration of the coupled system.

Under these assumptions, the equations for the auxiliary fields
can be solved analytically in terms of the scale factor. Using
$\dot f = aH f'$ where $f' \equiv df/da$, and neglecting corrections
of order $\dot H/H^2$, one finds
$\ddot{f} +3H \dot{f}  = H_c^2(a^2 f'' + 4a f')$.
Then,  Eqs.~\eqref{eq:zeta} - \eqref{eq:chi} yield
\begin{align}
& \xi(a) = A_\xi - \frac{B_\xi}{a^3},
\\
& \zeta (a) = A_\zeta - \frac{B_\zeta}{a^3} + \frac{M^2}{4H_c^2 a^{4}} ,
\\
& \varphi(a) = A_\varphi - \frac{B_\varphi}{a^3}
+ \Big( \frac{2A_\xi}{3H_c^2} - 8  \Big)\ln a
+ \frac{2 B_\xi}{3 H_c^2} \frac{\ln a}{a^3},
\\
& \chi(a) = A_\chi  - \frac{B_\chi}{a^3}
+ \frac{2A_\zeta}{3H_c^2} \ln a
+ \frac{2 B_\zeta}{3 H_c^2} \frac{\ln a}{a^3}
+ \frac{M^2}{8 H_c^4 a^4 }.
\end{align}
These solutions contain homogeneous modes, controlled by the
integration constants $A_i$ and $B_i$, and particular solutions
for which the curvature and radiation sectors serve as sources.
The auxiliary fields arise from the localization of the nonlocal
operators, whose inverse d'Alembertian is defined through a
prescribed Green function. Consequently, in the parent nonlocal
theory, the auxiliary fields are uniquely determined by the sources
$R$ and $F^2$, as expressed in Eqs.~\eqref{auxphchi} and
\eqref{aux_xize}. By contrast, the localized equations are
second-order differential equations and therefore admit additional
homogeneous solutions that are not fixed by these sources. Retaining
such homogeneous modes would introduce extra degrees of freedom
that are absent in the original nonlocal formulation, thereby
enlarging the solution space of the localized theory.
We therefore restrict our analysis to the source-induced branch
by setting $A_i = B_i = 0$.\footnote{
It is worth emphasizing that this choice removes the homogeneous 
modes only in the asymptotic quasi-de Sitter regime used to construct 
the initial data. The subsequent cosmological evolution is obtained 
by integrating the full system of equations. Therefore, in the 
nonlinear regime, maintaining an exact correspondence between the 
solutions of the local and nonlocal formulations is not guaranteed 
and may require a suitable tuning of the initial conditions and model 
parameters.}
With this simplification, the solutions reduce to
\begin{align}
\xi = 0,
\qquad \qquad
\zeta(a) =  \frac{M^2}{4H_c^2 a^{4}},
\qquad
\varphi(a) = - 8\ln a,
\qquad
\chi(a) =  \frac{M^2}{8H_c^4 a^{4}}.
\end{align}

The initial time is specified by choosing an initial value of the scale
factor $a_i$ during the contracting phase. In the
analysis in Ref.~\cite{BoRa24},  there was obtained the minimum
value of the scale factor given by
\begin{align}
a_m = a_0\, \exp\left\{ - \frac{1}{2\kappa M_P^2}\right\},
\label{eq:am}
\end{align}
where $a_0$ denotes the reference scale at which the asymptotic
contracting solution is normalized. Using the above value as a
reference, the initial conditions can then be imposed on
\begin{align}
a_i = \lambda a_m,
\end{align}
where $\lambda > 1$ is a parameter that determines how far
from the bounce the numerical integration begins. Since
$a_m < a_0 < a_i$, the integration starts deep in the asymptotic
contracting regime, where the analytical solutions above provide
reliable initial data.

To perform the numerical integration and obtain the complete
coupled solution of the system
\eqref{Friedmann_eq} - \eqref{eq:chi}, we regarded the scale
factor $a$ to be an independent variable in all equations and
combined the modified Friedmann equations, \eqref{Friedmann_eq}
and \eqref{00_eq}, to eliminate $\Lambda$. This procedure yields
the following differential equation for the Hubble parameter
\begin{align}
(1-2\kappa \zeta)aHH' + \frac{\kappa}{2}\bigg[a^2 H^2\chi'\xi'
+ 8aH^2 \zeta' + a^2 H^2\varphi'\zeta' - 2F^2
- \frac{1}{3}F^{2}\varphi\bigg] = 0.
\label{maineq}
\end{align}
This equation served as a basis for numerical calculations.

\begin{figure}[ht!]
	\centering
	\includegraphics[scale=.85]{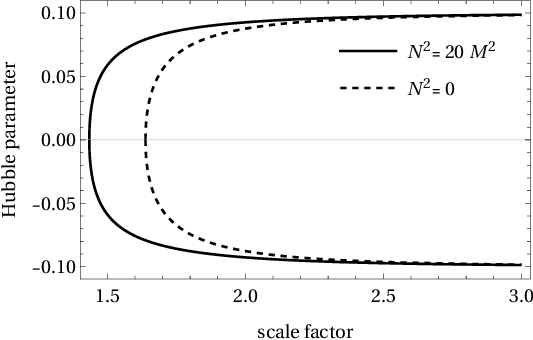}
	\includegraphics[scale=.85]{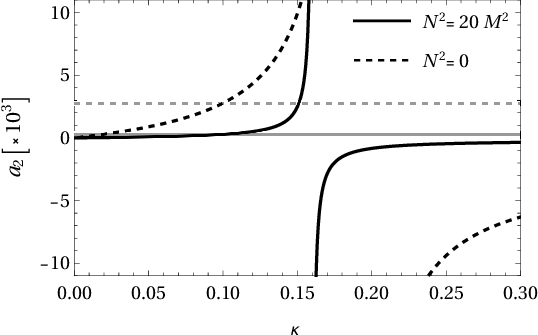}
	\vspace{-5mm}
\begin{quotation}
	\caption{
Numerical results obtained from initial conditions fixed during
the contracting phase, for $N^2=0$ and 
$N^2=20M^2$ (dashed line).
Left panel shows the Hubble parameter in Planck units, with $M^2=0.5$,
$\kappa=0.1$, $a_0=1$, and $\lambda=10^{4}$. The two branches
of the solution correspond to asymptotic quasi-de Sitter contracting
and expanding regimes, with asymptotic Hubble parameters
$H_c=-0.1$ and $H_c=0.1$, respectively. Right panel shows the
coefficient $a_2$ as a function of $\kappa$, obtained from
Eq.~\eqref{eq:a_2general}. The horizontal lines correspond to the
values of $a_2$ for $\kappa = 0.1$.}
\label{fig1}
\end{quotation}
\end{figure}

Fig.\;\ref{fig1} summarizes our numerical results. The left panel
shows the evolution of the Hubble parameter as a function of the
scale factor, obtained by stitching together the contracting ($H<0$)
and expanding ($H>0$) branches of the numerical solution for
two choices for the anomalous parameter, namely $N^{2}=0$ and
$N^{2}=20M^{2}$. The inclusion of the anomalous contribution
proportional to $N^{2}/(\varphi^{2}a^{4})$ shifts the bounce
toward smaller values
of the scale factor, from $a_b\simeq1.64$ for $N^{2}=0$ and
 $a_b\simeq 1.44$ for $N^{2}=20M^{2}$.
It is worth remembering that the definitions of the integration
constants $N$ and $M$ can be found in Appendix \ref{apdx}.

As the Universe contracts, the growth of the effective radiation
density drives a departure from the asymptotic quasi-de Sitter
regime, leading to a smooth transition through the bounce point
at $H=0$.
The right panel displays the coefficient $a_2$, defined in
Eq.~\eqref{eq:a_2general}, as a function of $\kappa$. The positivity
of $a_2$ throughout the explored parameter range shows that the local
bounce condition remains satisfied even in the presence of sizeable
anomalous corrections. In particular, increasing $N^{2}$ modifies the
quantitative properties of the bounce, such as the value of $a_b$,
but does not spoil the local condition $a_2>0$ required for a regular
nonsingular transition.
These results show that the covariant local formulation preserves the
semiclassical bounce mechanism, retaining the main physical features
of the solution reported in Ref.~\cite{BoRa24}.

It is worth noting that choosing the initial hypersurface too close to
the bounce leads to premature numerical stiffness due to the steep
scaling of the complete anomalous radiation. This reflects the fact
that the analytical auxiliary-field solutions provide only leading-order
asymptotic behavior of the contracting phase, whereas the subsequent
evolution is governed by the full coupled system. In practice, we
find that values of order $\lambda \sim 10^4$ are sufficient to
obtain a stable numerical evolution toward the nonsingular bounce.

The case $\beta<0$ is not compatible with the present construction.
Indeed, since $\kappa\propto\beta$, Eq.~\eqref{eq:am} implies that
$\beta>0$ leads to $a_m<a_0$, allowing the Universe to evolve from the
asymptotic contracting regime toward the bounce. Conversely,
$\beta<0$ yields $a_m>a_0$, reversing this hierarchy and rendering the
assumed contracting scenario inconsistent.

\subsubsection{Initial conditions in the bounce}
\label{sec3.2.2}

Contrary to the previous construction, we do not assume an asymptotic
contracting phase. Instead, we investigate whether the nonsingular
transition emerges as an intrinsic property of the coupled dynamical
system (\ref{Friedmann_eq}) - (\ref{eq:chi}).

According to the requirements derived in
Sec.~\ref{sec:bouce-conditions}, the first derivative of the
auxiliary fields can be taken vanishing at $t_b$. Then, we set the
conditions \eqref{eq:dfields0}, and
\begin{align}
\xi_b = \chi_b = \zeta_b = 0\,.
\end{align}
The value $\varphi_b=0$ is no longer admissible whenever $N^2\neq0$
because it would make $F^2$ singular. The local bounce condition
derived in Sec.~\ref{sec:bouce-conditions} provides a useful
restriction on the allowed values of $\varphi_b$. For $\zeta_b=0$
and vanishing first derivatives, one finds
\begin{align}
a_2 \,=\, \frac{\kappa}{4}
\bigg( \frac{M^{2}}{a_b^4} - \frac{N^{2}}{\varphi_b^2 a_b^4}\bigg)
\left( 2 + \frac{\varphi_b}{3} \right).
\end{align}
This relationship shows that the values of $\varphi$
corresponding to $a_2 > 0$ satisfy the inequality
\begin{align}
- \,6 \,< \, \varphi_b
\,<\, -\,\sqrt{\frac{N^2}{M^2}}\,,
\label{eq:neg_interval}
\end{align}
or, alternatively,
\begin{align}
\varphi_b \,>\, \sqrt{\frac{N^2}{M^2}}.
 \label{eq:pos_interval}
\end{align}
The first interval exists only when $\sqrt{N^{2}/M^{2}} < 6$,
as otherwise the admissible region reduces to the second interval.

As before, the numerical integration is performed after eliminating
$\Lambda$ by combining Eqs.~\eqref{00_eq} and \eqref{Friedmann_eq},
thereby obtaining a single evolution equation for $\dot H$.
Although both intervals above guarantee the existence of a local
minimum of the scale factor, the subsequent evolution governed by
Eqs.~\eqref{Friedmann_eq}--\eqref{eq:F2} exhibits qualitatively
distinct behaviors. For this reason, the two intervals are analyzed
as the two separate cases.

\begin{figure}[ht!]
\centering
\includegraphics[scale=.9]{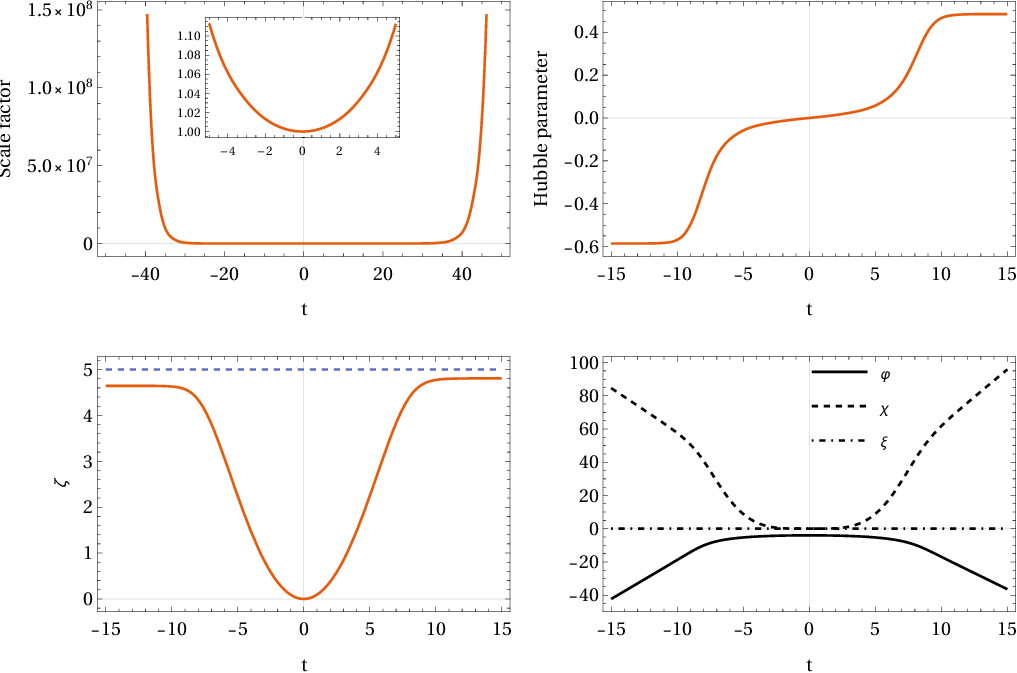}
\vspace{-5mm}
\begin{quotation}
\caption{ 
Plots of the scale factor, the Hubble parameter, and the auxiliary
fields as functions of the cosmic time (in Planck units) for
$\varphi_{b} = -4$,  $M^{2} = 0.5$, $\kappa = 0.1$, $a_b = 1$
and $N^{2} = 10 M^{2}$. Top panels correspond to $a(t)$ (left)
and $H(t)$ (right). The bottom left shows the evolution of the auxiliary
field $\zeta$ together with the stability threshold $1/(2\kappa)$
(dashed line). The bottom right illustrates the evolution of the
remaining auxiliary fields, namely $\xi$ (dot-dashed), $\chi$
(dashed), and $\varphi$ (solid).
}
\label{fig:2}
\end{quotation}
\end{figure}

\textit{i.} \
Negative interval \eqref{eq:neg_interval}:
Figure~\ref{fig:2} presents the numerical results for representative
values within this interval, where we choose $\varphi_b=-4$ and
$N^2=10M^2$. The upper-left panel shows the evolution of the
scale factor, which exhibits a smooth bouncing transition. The
corresponding evolution of the Hubble parameter is displayed in
the upper-right panel. The lower-left panel compares the evolution
of $\zeta(t)$ with the critical value $1/(2\kappa)$, showing that
the condition $\zeta(t)<1/(2\kappa)$ is satisfied throughout the
entire evolution. The lower-right panel displays the behavior of
the remaining auxiliary fields. Our numerical analysis yields
similar results for other parameter choices within the allowed
region of this interval.

\begin{figure}[ht!]
\centering
\includegraphics[scale=.9]{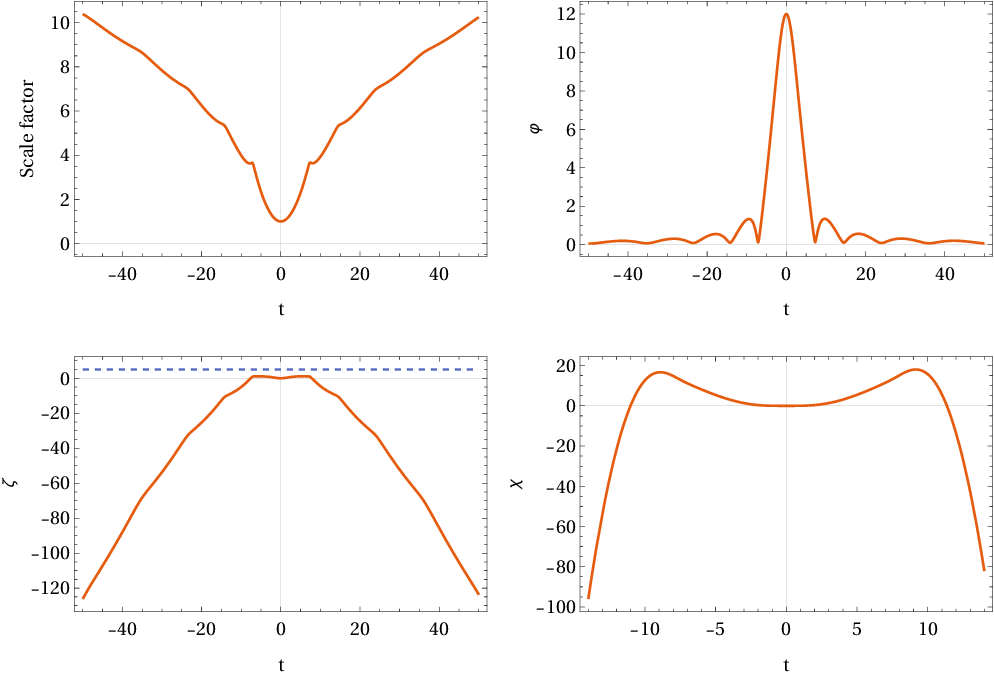}
\vspace{-5mm}
\begin{quotation}
\caption{
Plots of the scale factor, the Hubble parameter, and the auxiliary
fields as functions of the cosmic time (in Planck units) for
$M^{2} = 0.5$, $\kappa = 0.1$, $a_b = 1$, $N^{2} = 40 M^{2}$ and
$\varphi_{b} = 12$. Top panels: $a(t)$ (left) and $\varphi(t)$
(right); bottom left: evolution of the auxiliary field $\zeta$ together
with the stability threshold $1/(2\kappa)$ (dashed blue line); bottom
right: evolution of the field $\chi$.}
\label{fig:3}
\end{quotation}
\end{figure}

\textit{ii.} \
Positive interval \eqref{eq:pos_interval}: For stronger couplings,
$N^2\geq36M^2$, the negative interval ceases to exist, forcing the
system to start within the positive interval. Figure~\ref{fig:3}
summarizes the numerical results and illustrates the typical
behavior of the background variables in this regime. As can be
seen from the upper-left panel, the scale factor develops
oscillations and rapid transition structures that become increasingly
sharp for larger values of $N^2$. The field $\xi$ is not shown
because it remains identically zero throughout the evolution.
Likewise, $H(t)$ follows the same intricate pattern displayed
by the scale factor, alternating periods of expansion with brief
episodes of contraction.

We note that this behavior is not a consequence of numerical
instability, but rather a genuine dynamical signature of the model.
Starting from an initial condition with
$\varphi_b >\sqrt{N^{2}/M^{2}}$, the coupled equations
of motion drive the auxiliary scalar field $\varphi(t)$ toward
smaller values after the bounce.  When $\varphi(t)$ approaches
zero, the singular contribution $-N^2/(\varphi^2 a^4)$ grows rapidly
in the negative direction. This term acts as a steep repulsive
dynamical barrier, preventing the field from crossing the singular
point $\varphi=0$. Consequently, $\varphi(t)$ is driven back
toward larger positive values, and the process repeats itself
successively. The result is a regime of rapid oscillations. Since
the coupling term $F^2(t)$ acts directly as a source in the
acceleration equation \eqref{Friedmann_eq}, this mechanism
induces corresponding variations in the derivative of the Hubble
parameter. The scale factor $a(t)$ reflects this dynamics, giving
rise to the characteristic behavior observed in Fig.\;\ref{fig:3}.

All in all, the described numerical results reflect the fact that the 
covariant model has a much broader mathematical solution space than 
the non-covariant description. Only a particular branch of these 
solutions is expected to reproduce the effective dynamics encoded 
in Ref.~\cite{BoRa24}. Consequently, an exact quantitative 
correspondence between the two formulations requires a extremely 
precise fine-tuning choice of the auxiliary-field initial data and 
other parameters in the bounce. Outside this restricted branch, the 
new formulation naturally admits a wider class of nonsingular 
cosmological evolutions, while preserving the same underlying 
semiclassical mechanism responsible for the bounce.

\section{Conclusions and discussions}
\label{sec6}

As the main result of this work, we constructed a covariant local
representation of the low-energy version of anomaly-induced
effective action. The locality of effective action is provided by
taking a consistent low-energy limit and subsequently introducing
an appropriate set of auxiliary scalar fields. As a result, there
are no higher derivatives in the effective equations. The approach
we followed can be applied either to the pure gravity without
matter, to the theory with the background electromagnetic field
\cite{GianMott09}, to scalars \cite{AnoInt23}, or to other choices
of an external field or fields.

As a testing of the new approximate form of the anomaly-induced
effective action, we analyzed the cosmological consequences
for the homogeneous and isotropic Universe, which is initially in
the contracting phase and is not empty. Our new formulation
provided a closed system of coupled equations for the
Hubble parameter and four auxiliary scalar fields, allowing a
more comprehensive study of semiclassical bouncing solutions
within a manifestly covariant framework.
It is remarkable that we could reproduce the main qualitative 
features of the bounce solution obtained in \cite{BoRa24}, which 
was derived from the non-covariant (albeit complete, that is not 
restricted to the IR region) form of the anomaly-induced effective 
action of gravity and radiation.

Assuming an isotropic and homogeneous background, we first
establish analytically the local conditions for a regular bounce.
We then solve the full nonlinear system numerically and show that
these conditions indeed evolve into complete nonsingular bouncing
cosmology. To demonstrate that the bounce mechanism is not
an artifact, we construct two sets of initial data. The numerical
results obtained in both cases show that the nonsingular transition
is a robust feature of the model based on the low-energy
approximation.

Our results also reveal two qualitatively distinct dynamical regimes
associated with the auxiliary field $\varphi$.
The negative interval provides the cleanest and most tractable
scenario for the analytical study of linear cosmological perturbations.
On the other hand, the positive interval exhibits a dynamical behavior
characterized by rapid oscillations arising from the nonlinear coupling.
This oscillatory dynamics may produce characteristic signatures, such
as superimposed oscillations, in the primordial power spectrum of the
CMB anisotropies, and generate nontrivial primordial
non-Gaussianities, thus potentially establishing a direct connection
between the semiclassical corrections induced by the conformal
anomaly and future high-precision cosmological observations.

The further developments may be two-fold. First, it would be very
interesting to apply the new version of the anomaly-induced action
in other physical situations which agree with the sub-Planckian
approximation. On the other hand, it is possible to explore the
cosmological perturbations for the bounce model in the IR
domain, without risking to meet higher-derivative instabilities.
Finally, the approach based on the IR limit in the anomaly-induced
theory may be operational for the theories beyond QED, where the
stronger interactions will make the bounce model of \cite{BoRa24}
numerically consistent with the sub-Planckian regime.

\section*{Acknowledgments}

W.C.S. is grateful for the financial support from \textit{Conselho
Nacional de Desenvolvimento Cient\'{i}fico e Tecnol\'{o}gico}
(CNPq - Brazil) for the Postdoctoral Fellowship [PCI grant number
314125/2025-6].
The work of I.Sh. is partially supported by
CNPq under the grant 305122/2023-1.


\appendix
\section*{Appendix}
\addcontentsline{toc}{section}{Appendices}
\renewcommand{\thesubsection}{\Alph{subsection}}
\subsection{Derivation of the solution to Equation \eqref{eq:F2}}
\label{apdx}

In this Appendix, we present the detailed calculations for the
radiation term given in Eq.~\eqref{eq:F2}. In a homogeneous
and isotropic universe described by the FLRW metric,
Eq.~\eqref{FLRW_metric}, we can expand
\begin{align}
F^{2} = F_{\mu\nu}F^{\mu\nu}
= 2 g^{00}g^{ij}F_{i0}F_{j0} + g^{ik}g^{jl} F_{ij}F_{kl}
= -\frac{2}{a^2}\delta^{ij}F_{i}^{~0}F_{j}^{~0}
+ \frac{1}{a^{4}}\delta^{ik}\delta^{jl} F_{ij}F_{kl} \,.
\label{eq:F2apdx}
\end{align}

To determine the components
$F_{i}^{~0}$, we solve Eq.~\eqref{eq:inc-rad} in the form
\begin{align}
\nabla_{\nu}(\varphi F^{\mu\nu})
= \partial_{\nu}(\varphi F^{\mu\nu})
+ \Gamma^{\mu}_{\nu\alpha}\varphi F^{\alpha\nu}
+ \Gamma^{\nu}_{\nu\alpha}\varphi F^{\mu\alpha} = 0.
\end{align}
Since the Maxwell tensor $F^{\mu\nu}$ is antisymmetric whereas
the Christoffel symbols are symmetric in their lower indices, the
contraction $\Gamma^{\mu}_{\nu\alpha}\varphi F^{\alpha\nu}$
vanishes. Moreover, using the identity $\Gamma^{\nu}_{\nu\alpha}
= \partial_{\alpha}(\ln\sqrt{-g})$, the above equation reduces to
\begin{align}
\partial_\nu(\varphi F^{\mu\nu})
+ \partial_{\alpha}(\ln\sqrt{-g})(\varphi F^{\mu\alpha}) = 0\,.
\end{align}
All background quantities depend only on the cosmic time. Thus,
for the spatial components $\mu=i$, the spatial derivatives vanish
and the previous equation reduces to
\begin{align}
\frac{d}{dt}(\varphi F^{i0})
+ \frac{3}{a} \frac{d a}{dt}(\varphi F^{i0}) = 0,
\end{align}
where we have used $\sqrt{-g}=a^3$. Multiplying by $a^3$, the
left-hand side can be written as a total derivative,
$\frac{d}{dt}\big(a^{3}\varphi F^{i0}\big) = 0$, from which
it follows that
\begin{align}
F^{i0}(t) = \frac{N^i}{\varphi(t) a^{3}(t)},  \label{eq:F0i}
\end{align}
where $N^i$ are integration constants fixed by the 
initial conditions.  This implies 
\begin{align}
F_{i}^{~0}(t) = -\frac{\delta_{ij}N^{j}}{\varphi(t) a(t)}\,.
\end{align}

Next, to determine the spatial components $F_{ij}$, we employ
the Bianchi identity,
\begin{align}
\partial_{\alpha} F^{\beta\gamma}
+ \partial_\beta F_{\gamma\alpha}
+ \partial_\gamma F_{\alpha\beta} = 0,
\end{align}
which reduces to $\dot{F}_{ij} = 0$, hence
$F_{ij} =  M_{ij}$, where $M_{ij}$ is a constant antisymmetric matrix.

Substituting the obtained expressions into Eq.~\eqref{eq:F2apdx} yields
\begin{align}
F^2(t) = \frac{M^2}{a^{4}(t)}-\frac{N^2}{\varphi^{2}(t)a^{4}(t)},
\end{align}
where $N^{2} = 2 \de_{ij}N^i N^j$ and
$M^{2} = \delta^{ik}\delta^{jl}M_{ik}M_{jl}$
are integration constants.


\end{document}